\documentclass[aps,prl,superscriptaddress,twocolumn,final,showpacs,showkeys,nobibnotes,titlepage,twoside,10pt]{revtex4-1}

\usepackage{amsmath}
\usepackage{natbib}
\usepackage{color}
\usepackage{graphicx}
\usepackage{siunitx}						
\usepackage[utf8]{inputenc}
\newcommand{\delete}[1]{\textcolor{red}{}}
\newcommand{\myeq}{\!=\!}
\newcommand{\tauLJ}{\tau_\text{LJ}}
\newcommand{\Tc}{T_\text{c}}
\newcommand{\rc}{r_\text{c}}

\begin{document}

\title{ Probing the degree of heterogeneity within a shear band of a model glass}

\author{Muhammad Hassani}
\email[]{These authors contributed equally to this work.}

\author{Alexandra E. Lagogianni}
\email[]{These authors contributed equally to this work.}

\author{Fathollah Varnik}
\affiliation{ICAMS, Ruhr-Universität Bochum, Universitätstraße 150, 44780 Bochum, Germany}
\thanks{Corresponding author: fathollah.varnik@rub.de}

\date{\today}

\begin{abstract}
Recent experiments provide evidence for density variations along shear bands (SB) in metallic glasses with a length scale of a few hundreds nanometers. Via molecular dynamics simulations of a generic binary glass model, here we show that this is strongly correlated with variations of composition, coordination number, viscosity and heat generation. Individual shear events along the SB-path show a mean distance of a few nanometers, comparable to recent experimental findings on medium range order. The aforementioned variations result from these localized perturbations, mediated by elasticity.
\end{abstract}


\maketitle

\section{Introduction}

One of the most prominent manifestations of heterogeneity upon deformation of  amorphous materials, such as metallic glasses~\cite{Greer2013b,Maass2015b} is the shear-banding phenomenon. When metallic glasses are exposed to low temperatures and high mechanical load, shear bands (SBs) are formed via localization of strain in narrow regions of 5-100 nm thickness \cite{Donovan1981a,Pauly2009a}. A shear band, characterized by local increase of temperature \cite{Battezzati2008a,Miracle2011b}, excess free volume \cite{Spaepen1977a,Hassani2016a} and shear-induced softening \cite{Bei2006a,Wu2017a}, constitutes the main mechanism responsible for the limited plasticity of metallic glasses and the main cause of their catastrophic failure at room temperature \cite{Greer2013b}. The precursor of shear bands in metallic glasses is the appearance of some regions that undergo local yielding, the so-called shear transformation zones (STZs) \cite{Argon1979,Demetriou2006,Hassani2018a}, which consist of 10$\sim$100 atoms that display large non-affine atomic displacements and geometrically unfavored motifs (GUMs) \cite{Ding2014,Lagogianni2009b}. The displacement and strain field around a STZ closely resembles that of an Eshleby inclusion \cite{Eshelby1957a,Dasgupta2012c,Puosi2014b,Hassani2018a,Hassani2018b} and in this description the shear band comes as the result of correlated and aligned quadrupoles \cite{Hieronymus-Schmidt2017,Sopu2017,Dasgupta2012c}. 

Shear bands affect their vicinity within a range of up to hundreds of $\mu$m~\cite{Schmidt2015a,Rosner2014a,Maass2014a}, inducing structural heterogeneity and fluctuations of local mechanical properties along the SB-direction \cite{Liu2018a,Tsai2017a,Wagner2011a,Tonnies2015a}.~The local density within the shear band varies also spatially~\cite{Gross2018b,Mandal2012a} and is accompanied by  deflections of the SB-path, with respect to its propagation direction \cite{Schmidt2015a,Rosner2014a}. 

These observations suggest that shear band features cannot be described by average values but instead a position dependent analysis is needed to characterize the gradient of heterogeneity within a shear band. Even though computer simulations provide useful insight of phenomena and length scales that experiments can not always resolve, in this case the complex nature of the shear bands~\cite{Hassani2016a} and the large length scales associated with local density variations~\cite{Hieronymus-Schmidt2017} impeded a detailed quantitative analysis of this issue via computer simulations in 3D. Consequently, the origin of spatially varying patterns and the strong position dependent nature that properties exhibit along a shear band remains still an open question.

Here, we probe, via molecular dynamics simulations, the spatial variations of density and provide, for the first time, direct evidence for its correlations with coordination number, composition, excess free volume, plastic activity, local viscosity and energy generation rate within and along a shear band. We observe that every single quantity is closely connected to density, displaying variations on a similarly long length scale. In contrast, spatial arrangement of quadrupolar shear transformation events occur on a significantly shorter length scale of a few nanometers. This supports the idea that STZs are localized events~\cite{Argon1979,Sopu2017} which, when occurring in an elastic medium, can trigger long range perturbations~\cite{Hassani2018b}. At the same time, continuum mechanics models which assume a periodic alignment of quadrupolar stress-field perturbations need to be modified in order to account for this separation of length scales~\cite{Hieronymus-Schmidt2017}. Interestingly, the average STZ-distance agrees well with recent experimental reports on medium range order~\cite{Hilke2019}. This highlights further the close connection between local structural features and the self-organization of shear transformation events~\cite{Cao2009,Sopu2017}.

\begin{figure*}
	\begin{center}
		\includegraphics[width=15cm]{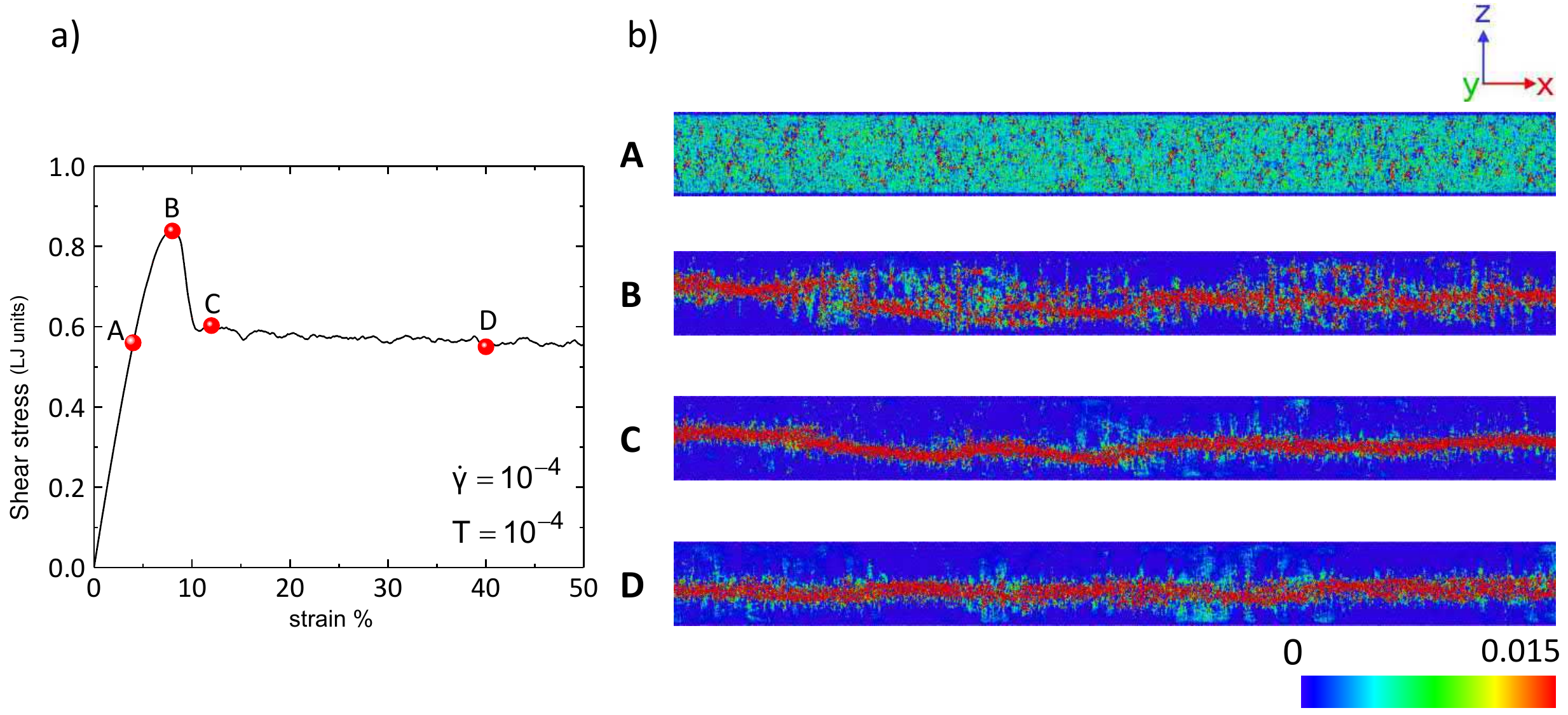}
		\caption{a) System averaged shear stress versus overall strain of the sheared glass in the athermal limit. Capital letters (A,B,C,D) indicate stress states for which the strain field is shown in the right panel. b) Color coded atomic strain in the shear ($xz$) plane for different stress states depicted as full red spheres in the left panel. Strain is evaluated using particle displacements within finite time intervals corresponding to 1\% global deformation. It contains both affine and non-affine contributions.}
		\label{fig:stress-vs-strain}
	\end{center}
\end{figure*}

A generic binary Lennard-Jones (LJ) glass former~\cite{Kob1994a} is used (Supplemental Material). Five statistically independent configurations are prepared. Each "sample" contains $N\approx{2.5}$ millions particles in a thin slab with dimensions $L_{x}\times L_{y}\times L_{z}=2000 \times 10 \times 100$ (reduced LJ units). The $L_x$ chosen here exceeds any earlier computationally-resolved scale in 3D, even though such orders of magnitude for spatial variations within shear band are suggested by several experimental works \cite{Liu2018a,Tsai2017a,Wagner2011a,Tonnies2015a}. Starting from an equilibrated liquid at a temperature of $T=2$ (the mode coupling critical temperature of the model is $\Tc \approx 0.43$ \cite{Kob1995a}), the system is quenched to a temperature of $T = 10^{-4}$ close to the athermal limit~\cite{Hassani2016a}. Simple shear is then imposed with a rate of $\dot{\gamma} =10^{-4}$ by relative motion of the two parallel walls along the $x$-direction \cite{Varnik2003a}. The walls correspond to two frozen layers, each of three particle diameters thickness, and are separated by a distance of $L_z=100$. Wall particles in this set of simulations have no thermal motion but move all together with a constant velocity of $\pm U_\text{wall}$ along the $x$-direction. Periodic boundary conditions are applied in the $x$ and $y$ directions. Using this protocol, we observe the formation of a single and stable system-spanning shear band without the need of any notch or stress concentrator. In order to maximize the overall size of the $xz$-plane while keeping the computational cost reasonable, the $y$ dimension of the box is set to $L_y=10$. This length is larger than the interaction cut-off length and the decay length of pair correlation function. Moreover, in the athermal limit considered here, finite size effects on dynamics play a sub-dominant role~\cite{Varnik2002c}. The systems is sheared up to 50\% overall strain. All the simulations reported here are performed using LAMMPS~\cite{Plimpton1995b} while the 3D visualization and the color coding is done by the OVITO software \cite{Stukowski2010a}.

The deformed glassy system displays a typical stress-strain response (Fig.~1a) with a stress overshoot, which depends on the imposed strain rate \cite{Varnik2004a}, followed by a shear softening region until a quasi-steady state is reached that extends up to the largest strain investigated ($\gamma_\text{max}=50\%$). The formation and the propagation of the shear band, demonstrated as high amounts of localized shear strain in a narrow region, is investigated via the calculation of the atomic strain from the infinitesimal Cauchy strain tensor, given as $\epsilon_{i,xz}=\frac{1}{2}\Big(\frac{\partial u_{i,x}}{\partial z}+\frac{\partial u_{i,z}}{\partial x} \Big) $, in a coarse-grained scheme \cite{Goldenberg2007a}. Here, $u_{i,\alpha}$ stands for the (coarse-grained) displacement of the $i$-th particle in the $\alpha=x,z$ direction, calculated  within a strain interval of $\delta \gamma = 1\% $.
 
Prior to yielding (point A in Fig.~\ref{fig:stress-vs-strain}a), the atomic strain is rather homogeneously distributed (image A in Fig.~\ref{fig:stress-vs-strain}b).  Progressively, as strain increases towards yielding, small isolated regions with accumulated atomic strain appear along the $x$-axis (B) which later coalesce into a system-spanning shear band (C and D) with a wavy character along the SB-propagation direction. This is in remarkable agreement with recent experimental observations \cite{Rosner2014a,Schmidt2015a}.

\begin{figure}
	\begin{center}
(a)		\includegraphics[width=0.9\columnwidth]{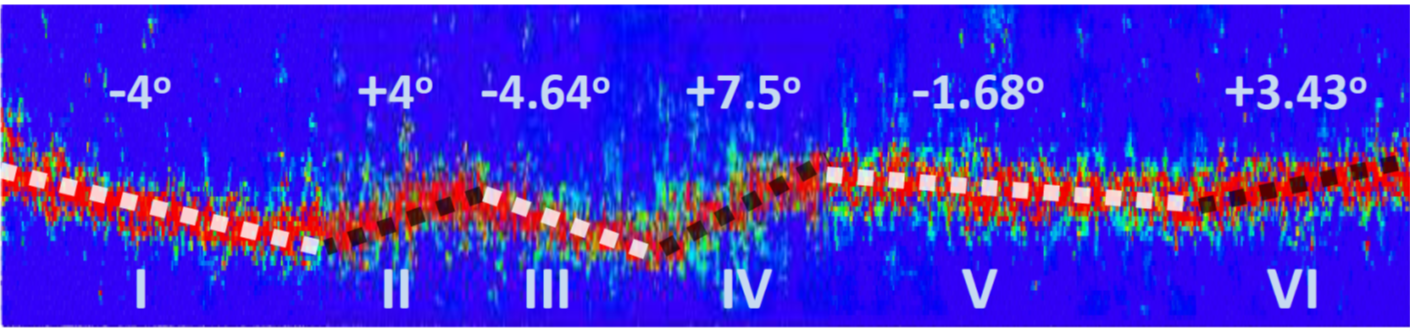}
(b)		\includegraphics[width=0.9\columnwidth]{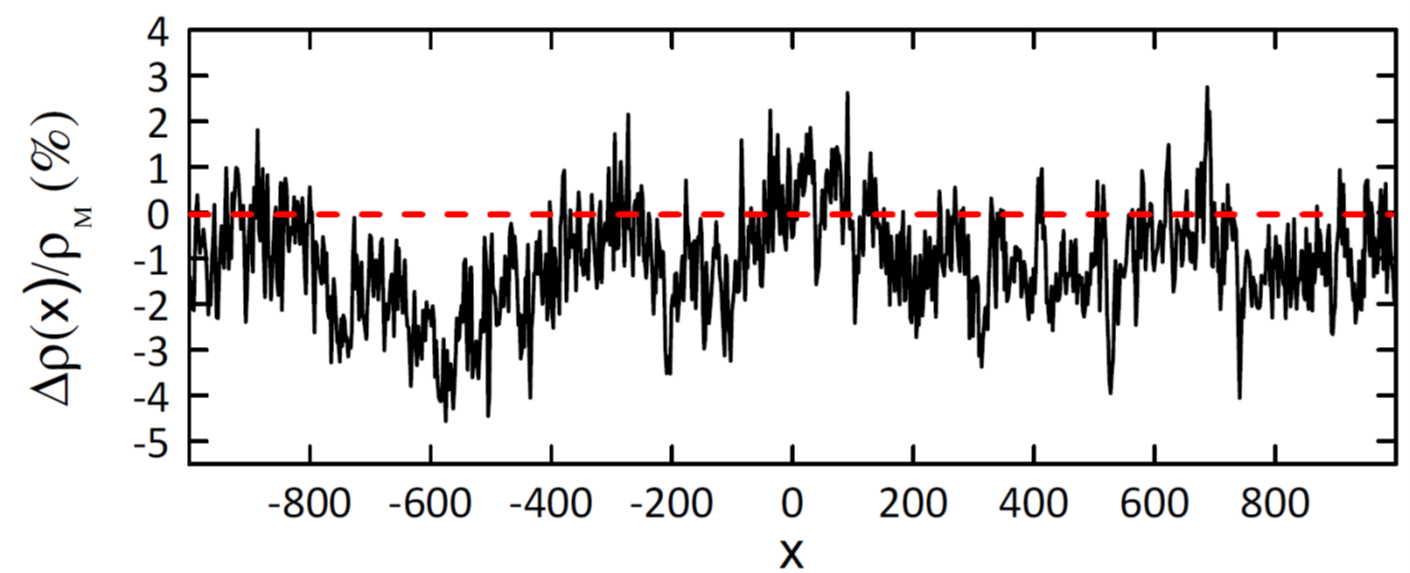}
		\caption{(a) A typical zig-zag type path of the shear band across the sample. White (black) dashed lines correspond to descending (ascending) segments. Note that, the aspect ration is not 1:1 but, to better highlight the wavy character of the SB-path, the image is vertically enlarged. (b) Spatial variations of the relative density differences, $\frac{\Delta\rho(x)}{\rho_\text{M}}$, along the SB-path.
		}
	\label{fig:SB-path}
	\end{center}
\end{figure}

The local deflections are quantified by binning the shear band path along the SB-propagation direction with a bin-width of two particle diameter. A geometric center (centroid) is assigned to each bin, calculated by the Cartesian coordinates of its $N_{\text{at}}$ constituent particles, $\vec{r} =  \frac{1}{N_\text{at}} \sum_{i=1}^{N_\text{at}} \vec{r}_i$ and averaged over sequential snapshots in the strain range $\gamma$=10$-$30 $\%$. The thus obtained centroids form a "chain" that represents geometrically the SB-path. For further analysis, an effective angle with respect to the $x$-axis is also assigned to each SB-bin. Sequential bins with negative or positive slope define a larger segment to which an average deflection angle is assigned. In agreement with experimental observations \cite{Schmidt2015a}, this analysis reveals alternating descending and ascending segments, labeled here by Latin numbers and highlighted as white and black dashed lines (Fig.~\ref{fig:SB-path}a). It has to be mentioned that this is a simplified representation of the shear band, where the segments are displayed as straight lines. The actual segments, however, are slightly curved (see images C and D in Fig.~\ref{fig:stress-vs-strain}b).

\begin{figure}
	\begin{center}
		\includegraphics[width=8cm]{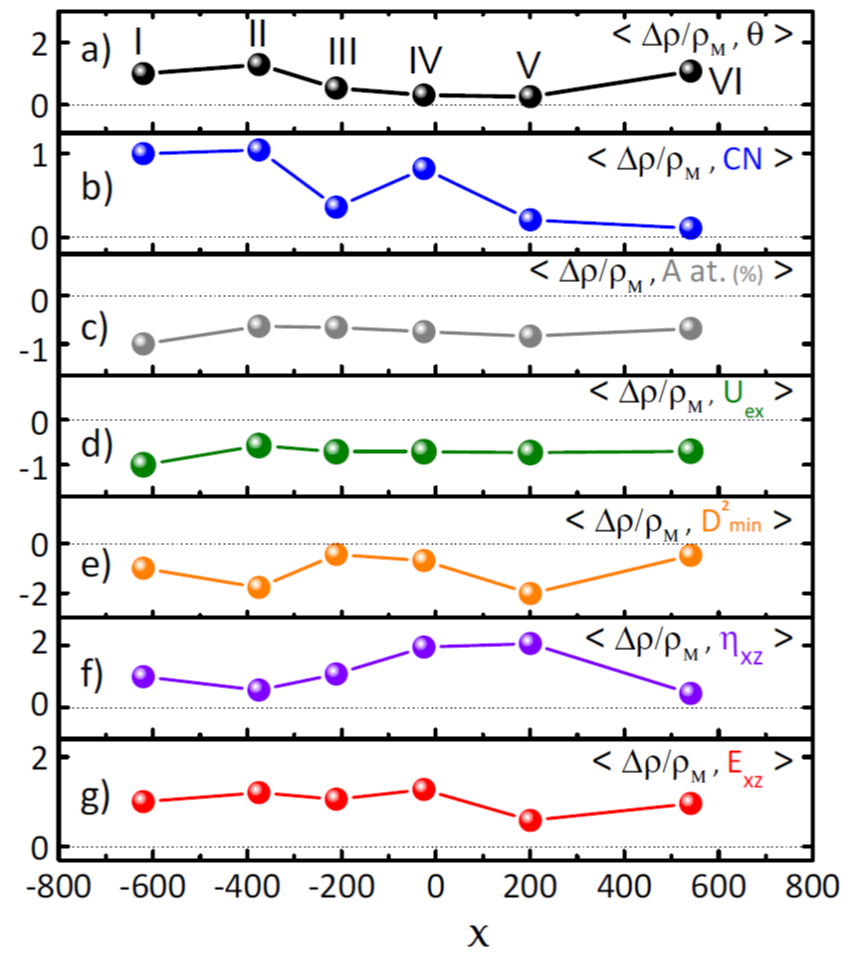}
\caption{Spatial correlations of the relative density difference within and outside shear band, $\Delta \rho/\rho_\text{M}$, with (a) the deflection angles, (b) the coordination number, (c) the composition/percentage ($\%$) of A particles in $A_xB_{100-x}$, (d) the excess volume $v_{i,\text{excess}}=v_{i,\text{voro}}-V/N$ \cite{Hassani2016a}, where $v_{i,\text{voro}}$ is the Voronoi volume  \cite{Voronoi1909a} and $V$ and $N$ are the total volume and number of particles in the system (e) the $D^2_\text{min}$ parameter \cite{Falk1998b}, (f) the dynamic viscosity $\eta=\sigma_{xz}/\dot{\gamma}$ ($\sigma_{xz}$=the $xz$-component of the atomic stress tensor, $\dot{\gamma}$=the strain rate in the $xz$-plane) and (g) energy dissipation, $E=\eta \dot{\gamma}^2$. All the data shown here are rescaled by the absolute value of the first (leftmost) data point,i.e., $f \to \hat f \equiv f/|f(-620)|$. This serves to better highlight whether a given quantity is correlated ($\hat f >0$) or anti-correlated ($\hat f <0$) with density. The fluctuations of each quantity are shown separately in the Supplementary Material and their respective amplitudes, characterized via standard deviations, are $4.9^\circ$ (inclination angle, $\theta$),  $0.14$ (coordination number, CN), 2\% (percentage of A-particles), $0.014$ (excess volume, $v_\text{ex}$), $0.14$ (plastic activity, $D^2_\text{min}$),  $117$ (viscosity, $\eta$) and $2.6\times 10^{-4}$ (viscous energy production, $E$).}
\label{fig:density-strain-corr}

\end{center}
\end{figure}

The local density changes inside and along the shear band path are spatially resolved by first introducing an atomic density, $\rho_i =\frac{1}{\Delta{V}}\sum_{j=1}^{N(i)}H(\rc-|\vec{r_i}-\vec{r_j}|)$, and then evaluating its averages within each domain of interest. Here, $H$ is the Heaviside step function and $N(i)$ is the number of particles enclosed in a sphere of radius $\rc$ (volume $\Delta{V}$) around the $i$-th particle ($\rc=$second minimum of the radial distribution function). The relative difference of the atomic density, averaged within each segment of the shear band, and its counterpart in the matrix, \(\frac{\Delta \rho(x) }{\rho_\text{M}} \equiv \frac{ \rho_\text{SB}(x)-\rho_\text{M} } {\rho_\text{M}}\) (Supplemental Material) displays a strong position-dependence and a wavy pattern that apparently averages out to a negative number indicative of a lower density within the shear band as compared to the matrix~\cite{Hieronymus-Schmidt2017}.

Density fluctuations show significant correlations with other physical quantities along the shear band. For example, we find that the density variations exhibit a positive correlation with the deflection angle (Fig.~\ref{fig:density-strain-corr}a) in agreement with experimental findings~\cite{Rosner2014a,Schmidt2015a}. Going beyond experiments, denser regions are found to have a higher average coordination number CN (Fig.~\ref{fig:density-strain-corr}b) and a lower percentage of large (A) particles (Fig.~\ref{fig:density-strain-corr}c). Since in general more unoccupied volume is available between larger spheres than among smaller ones, a region less rich in A-particles impedes the creation of excess free volume (Fig.~\ref{fig:density-strain-corr}d), and therefore decreases the possibility of this region to undergo a non-affine deformation (Fig.~\ref{fig:density-strain-corr}e), the latter quantified by the $D^2_\text{min}$ parameter \cite{Falk1998b}. In full agreement with this observation, we also find that the dynamic viscosity, that defines the local resistance of the system to plastic flow, is larger in regions of higher density (Fig.~\ref{fig:density-strain-corr}f). And finally, more viscous regions coincide with higher amounts of locally dissipated energy (Fig.~\ref{fig:density-strain-corr}g).

\begin{figure*}
\includegraphics[width=16cm]{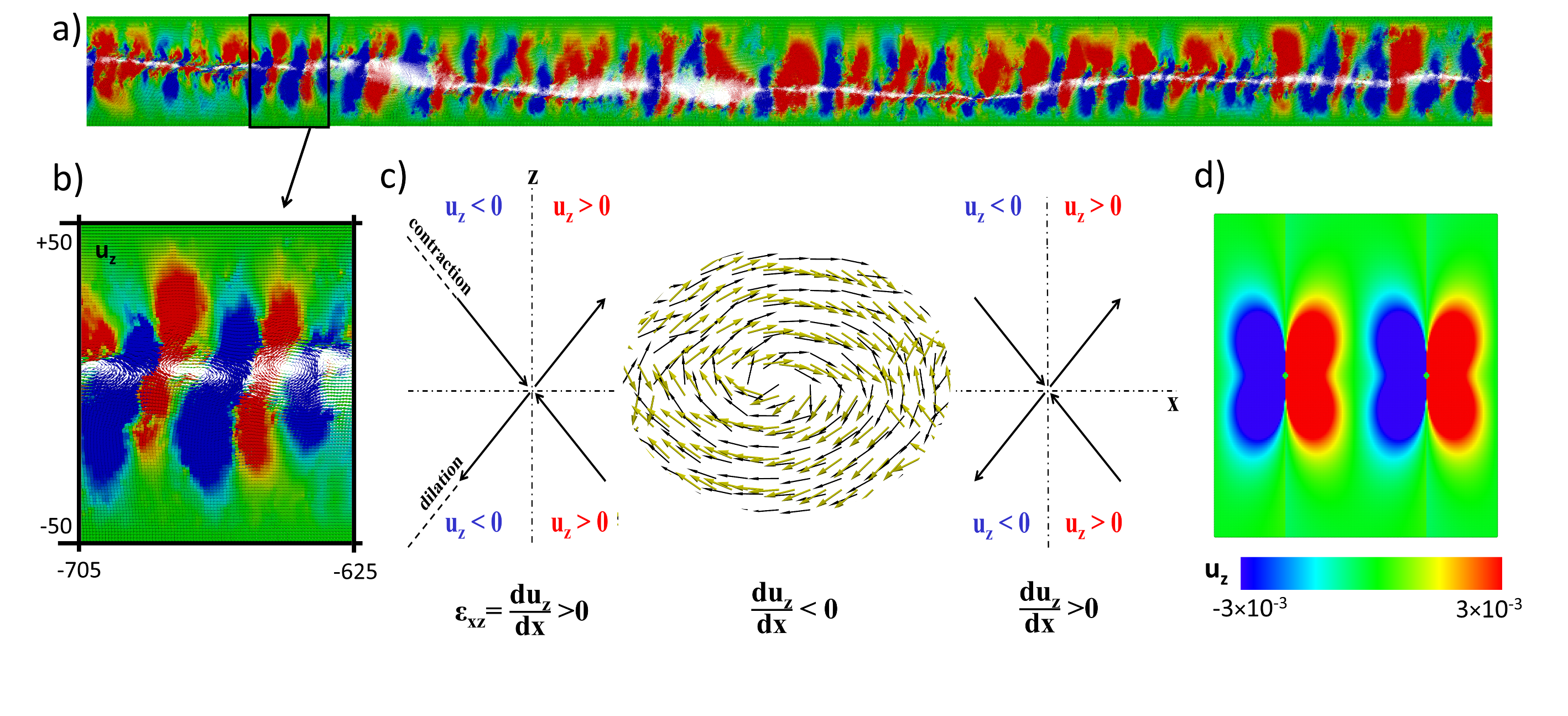}
\caption{(a) Scaled displacement vectors, $\vec u$, colored by its non-affine component, $u_z$. (b) The zoom of (a) around two adjacent STZs. (c) A further zoom into the region between the two STZs from (b) showing the vortex structure of the displacement vectors (black color). The vortex in gold is the continuum mechanics solution for the displacement vector field generated by two adjacent localized shear perturbations (color coded in panel d). The schematic drawing of the contraction and dilation axes serves to highlight the alternating sign of $\varepsilon_{xz}$ between two adjacent STZs (see Supplemental Material, Fig. S1).}
\label{fig:quadrupoles}
\end{figure*} 
\begin{figure}[h]
	\includegraphics[width=8cm]{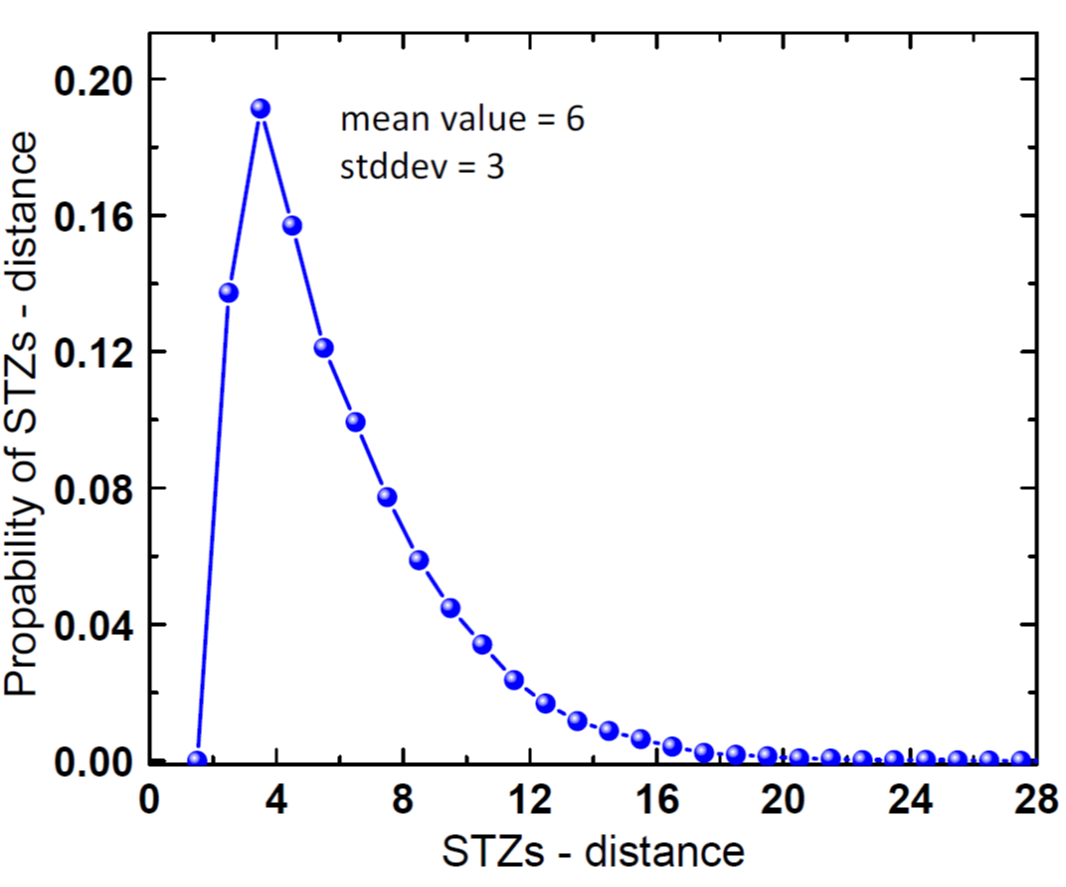}
	\caption{Distribution of the distance between two adjacent STZs along the shear band (reduced LJ units). Note that there are hardly two STZs with a distance above 20 particle diameters. This must be contrasted to the length scale of density variations, which span over hundreds of particle diameters (Fig.~\ref{fig:SB-path}).}
	\label{fig:histogram}
\end{figure}

It has been proposed recently that the variations of density within shear band originate from an alignment of quadrupolar stress-field perturbations~\cite{Hieronymus-Schmidt2017}. To address this issue, we build upon the analogy between the displacement field generated by a STZ and the corresponding continuum mechanics solution for a localized shear perturbation~\cite{Illing2016,Maier2017,Hassani2018a,Hassani2018b}. An analysis of the non-affine displacement field reveals a sequence of STZs along the entire shear band path (Fig.~\ref{fig:quadrupoles}a) with the same characteristics as the field generated by adjacent localized shear deformations in an isotropic elastic medium (Fig.~\ref{fig:quadrupoles}b-d). We thus identify adjacent STZs and find a skew-symmetric distribution of their distance with a mean value of roughly six particle diameters~(Fig.~\ref{fig:histogram}). A comparison to Fig.~\ref{fig:SB-path}b reveals that this value is by roughly two orders of magnitude smaller than the length scale associated with density variations. For a comparison with experimental data, we choose the size of a particle to be a few {\AA}ngstrom and find the same scale separation between the wave length of density variations and the distance between adjacent STZs (stress-concentrators).

In summary, the following picture emerges from our findings. (i) Strong density variations occur along shear bands with a wave length comparable to the system size; (ii) this is accompanied by gradients of composition both in longitudinal (along SB-path) and transverse (SB-matrix) directions; (iii) while plastic activity expectedly decreases with increasing density along the SB-path, the local heat generation rate is enhanced; (iv) quadrupolar shear transformation events are arranged on a rather short length scale, well separated from fluctuations of density. The agreement between density modulations in our simple glass model and experiments on bulk metallic glasses is remarkable~\cite{Maass2014a,Hieronymus-Schmidt2017} and underlines the generic character of this phenomenon. While tracking compositional changes in experiments is not an easy task, our finding motivates such experiments by showing that strong composition gradients may occur both between the matrix and shear band as well as along the SB-path. A possibility here would be to use radioactive tracers~\cite{Gaertner2019} and survey their spatial distribution during different stages of deformation. Energy generation within shear bands may lead to  local softening/rejuvenation effects and is of great importance both from practical and fundamental perspective. A thorough investigation of this issue thus provides an important topic for current research. Last but not least, the fact that shear transformation events are scale separated from the aforementioned variations of density and other important quantities underlines the deep connection between elasticity of the medium, which propagates signals on macroscopic lengths and localized perturbations, which excite eigenmodes of the sample. Simple continuum mechanics models, which lead to a correspondence between length scales of perturbation and response within the shear band need to be modified accordingly.

\section {Acknowledgments}
Fruitful discussions with A.~Zaccone are acknowledged.
This study has been financially supported by the German Research Foundation (DFG) under the project numbers VA205/16-2 and VA 205/18-1.
ICAMS acknowledges funding from its industrial sponsors, the state of North-Rhine Westphalia and the European Commission in the framework of the European Regional Development Fund (ERDF).

\newpage \clearpage
\newpage \clearpage

\setcounter{figure}{0}
\renewcommand\thefigure{S\arabic{figure}}

{\bf \Large Supplemental Material}\\

{\bf Molecular Dynamics Simulations}\\
The model consists of a binary mixture of 80\% large (A) and 20\% small (B) particles, interacting via the Lennard-Jones potential,
\begin{equation}
	U_\text{LJ}(r_{\alpha\beta}) = 4\epsilon_{\alpha\beta} 
	\Big[ 
	\Big( \frac{d_{\alpha\beta}}{r_{\alpha\beta}}\Big)^{12}-\frac{d_{\alpha\beta}}{r_{\alpha\beta}}\Big)^{6} 
	\Big]
\end{equation}
with $\alpha,\beta \myeq {\mathrm{A,B}}$, $\epsilon_{\mathrm{AB}}\myeq 
1.5\epsilon_{\mathrm{AA}}$, $\epsilon_{\mathrm{BB}}\myeq 
0.5\epsilon_{\mathrm{AA}}$, $d_{\mathrm{AB}}\myeq 0.8d_{\mathrm{AA}}$, 
$d_{\mathrm{BB}}\myeq 0.88d_{\mathrm{AA}}$ and $m_{\mathrm{B}}\myeq 
m_{\mathrm{A}}$. In order to enhance computational efficiency, the potential is 
truncated at twice the minimum position of the LJ potential, $r_{\text{c},\alpha\beta} \myeq  2.245 
d_{\alpha\beta}$. The parameters $\epsilon_{\mathrm{AA}}$, $d_{\mathrm{AA}}$ and 
$m_{\mathrm{A}}$ define the units of energy, length and mass, respectively.  The unit of time is a combination of these units, $\tauLJ \myeq d_{\mathrm{AA}}\sqrt{m_{\mathrm{A}} / \epsilon_{\mathrm{AA}}}$. The total number density is $\rho=\rho_\text{A}+\rho_\text{B}=1.2$. The simulation box contains $N\approx{2.5}$ millions particles and has a slab geometry with dimensions $L_{x}\times L_{y}\times L_{z}=2000 \times 10 \times 100$. Numerical time integration of Newton's equations of motion is done by the velocity-Verlet algorithm with an integration step of $dt=0.008$.

To evaluate density and other quantities within and outside the shear band, particles that belong to the shear band are distinguished from those of the matrix based on a strain criterion: A particle, $i$, is assigned to the shear band if its associated strain exceeds a threshold, i.e, if $\epsilon_{i, xz} \geq \epsilon_\text{cut-off}=0.1$ and to the matrix otherwise. We have checked that results reported in this work are not sensitive to the exact numerical value of this threshold, provided that it is close to the yield strain.

{\bf Correlations of density with other properties along the shear band path}
The correlation of the normalized density variations, here denoted as $F(x)$, with each of the other quantities, $G(x)$, is calculated as follows: 
$C(x)=\frac{1}{N_\text{s}} \sum\limits_{t = 1}^{N_\text{s}} \left< F(x) G(x) \right>$, where $F(x)=\frac{\Delta \rho(x)}{\rho_\text{M}}-\left<\frac{\Delta\rho(x)}{\rho_\text{M}}\right>$,  $G(x)=g(x)-\left<g(x)\right>$ and $N_\text{s}=21$ is the number of snapshots over which we have sampled the statistics. 

{\bf Two adjacent pre-sheared inclusions}\\
The displacement field around an inclusion with a traceless eigenstrain inside a homogeneous elastic medium, $\mathbf{\epsilon}^*$, is provided via analytical solution of Navier-Lam\'e equation in~\cite{Dasgupta2013}. Based on this solution, by assuming two eigenvectors lying in $xz$-plane with 45 degree angle to the horizontal axis, $\vec{n}=(\sqrt{2}/2,0,\sqrt{2}/2)$ and $\vec{k}=(\sqrt{2}/2,0,-\sqrt{2}/2)$, and their corresponding eigenvalues as $\lambda_n = \epsilon^*$ and $\lambda_k = -\epsilon^*$, the displacement field around the inclusion reads:
\begin{eqnarray}
	u_{x}(r,\theta) &=& \frac{a^3}{3}\frac{\epsilon^*}{(1-\upsilon)r^2}\left[ (1-2\upsilon) + \frac{3a^2}{5r^2}\right]\sin(\theta) \nonumber\\ 
	&+&\frac{a^3\epsilon^*}{2(1-\upsilon)r^2}\left[1 - \frac{a^2}{r^2}\right] 2\cos^2(\theta)\sin(\theta),
	\label{eq:specCase-Ux}
\end{eqnarray}	
and
\begin{eqnarray}
	u_{z} (r,\theta) &=& \frac{a^3}{3}\frac{\epsilon^*}{(1-\upsilon)r^2}\left[ (1-2\upsilon) + \frac{3a^2}{5r^2}\right]\cos(\theta) \nonumber \\ 
	&+& \frac{\epsilon^*a^3}{2(1-\upsilon)r^2}\left[1 - \frac{a^2}{r^2}\right] 2\sin^2(\theta)\cos(\theta).
	\label{eq:specCase-Uz}
\end{eqnarray}	
Here $\vec{r}$ also lies in $xz$-plane as $\vec{r}=r\left(\cos(\theta),0,\sin(\theta) \right)$, $a$ is the radius of inclusion and $\upsilon$ represents the Poisson ratio of the elastic medium.

\begin{figure*}
	\begin{center}
		\includegraphics[width=1.9\columnwidth]{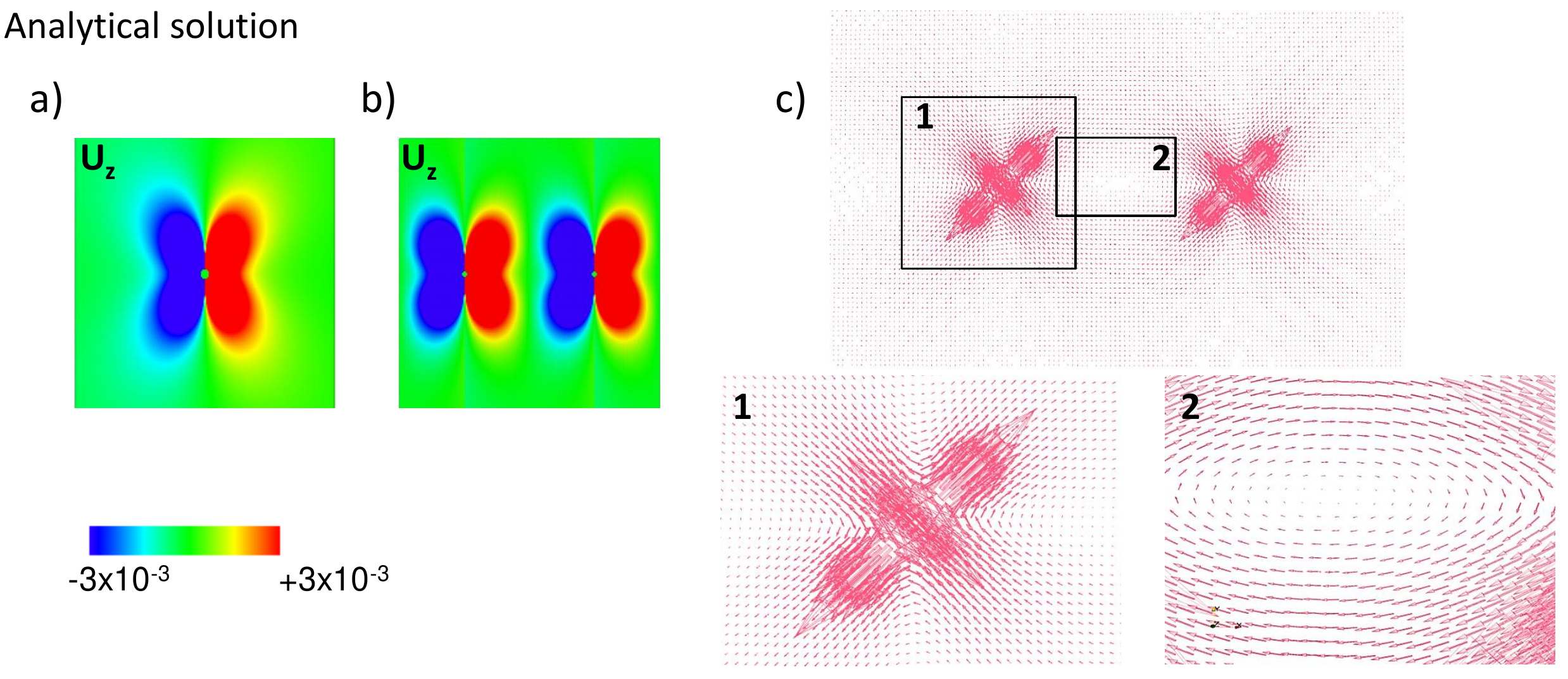}
		\caption{Continuum mechanics solution for the displacement field generated by (a) a single and (b) a pair of localized shear perturbations. Panels (a) and (b) show only the color-coded $z$-component of displacements, $u_{z}$. Panel (c) shows the displacement vector field corresponding to a pair of localized shear-events. The numbered images zoom to sections of the panel (c) and serve to highlights the vortex structure of $\vec u$.}
		\label{fig:figure-power1}
	\end{center}
\end{figure*}

To determine the displacement field around two adjacent pre-sheared inclusions, considering the linearity of the Navier-Lam\'e equation, the displacement field in Eqs.~\eqref{eq:specCase-Ux}~and~\eqref{eq:specCase-Uz} can be superposed with an offset in the reference frame for the position vector $\vec{r}$. According to this picture, $u_z$ around the two inclusions is given in Fig.~\ref{fig:figure-power1}. Interestingly, the displacement vectors between the inclusions also display a vortex-like pattern similar to observation in SB.

\begin{figure*}
	\begin{center}
		\includegraphics[width=1.3\columnwidth]{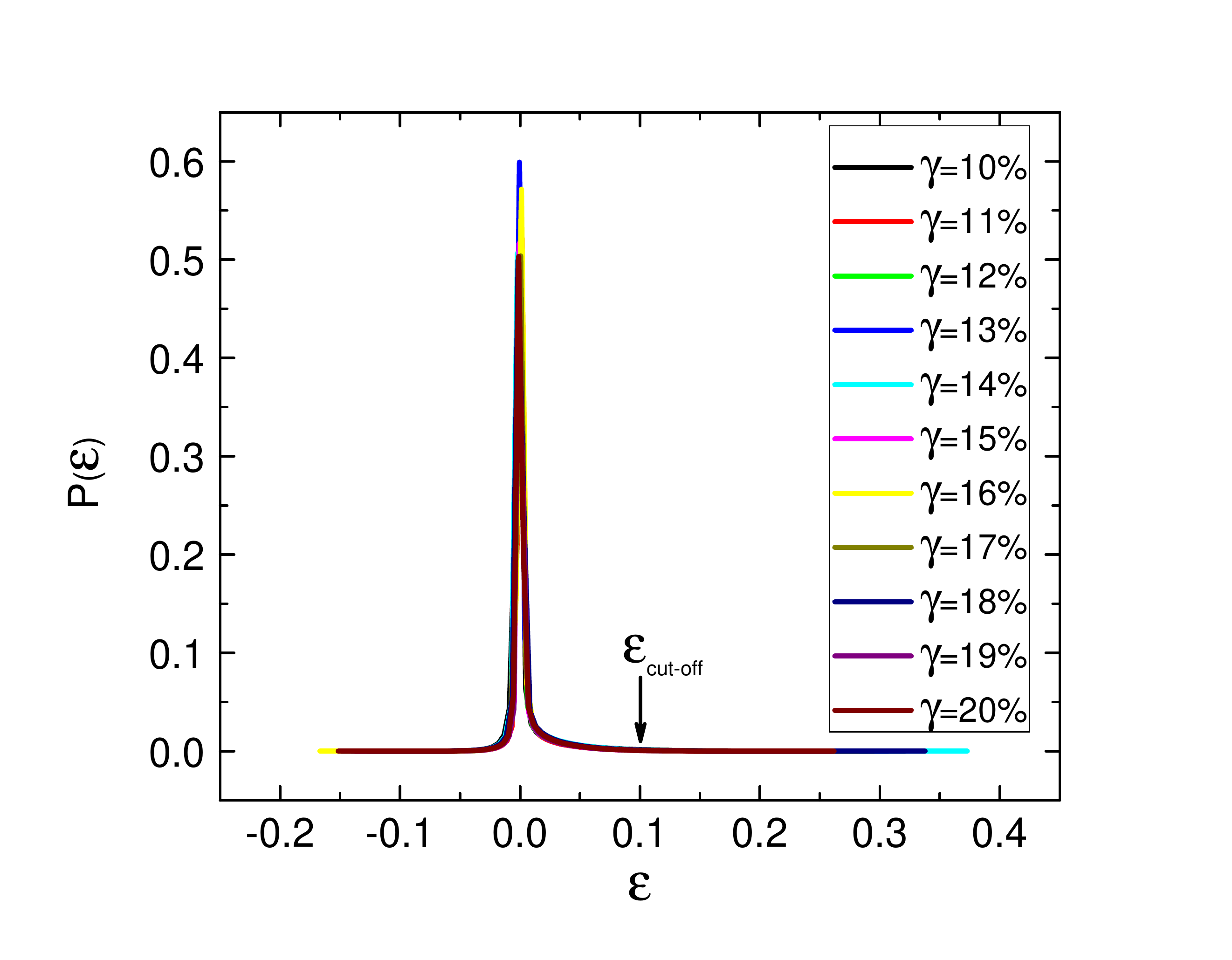}
		\caption{Histograms of the atomic strain, evaluated within 1\% intervals of the externally imposed shear deformation. Different curves correspond to different measurement times, or, equivalently, to overall strains, $\gamma =
			t\dot\gamma$. The vertical arrow marks the threshold strain, $\epsilon_\text{cut-off}$, used to distinguish shear band particles from those that belong to the matrix.}
		\label{fig:exz-histogram}
	\end{center}
\end{figure*}

\begin{figure*}
	\begin{center}
		\includegraphics[width=1.4\columnwidth]{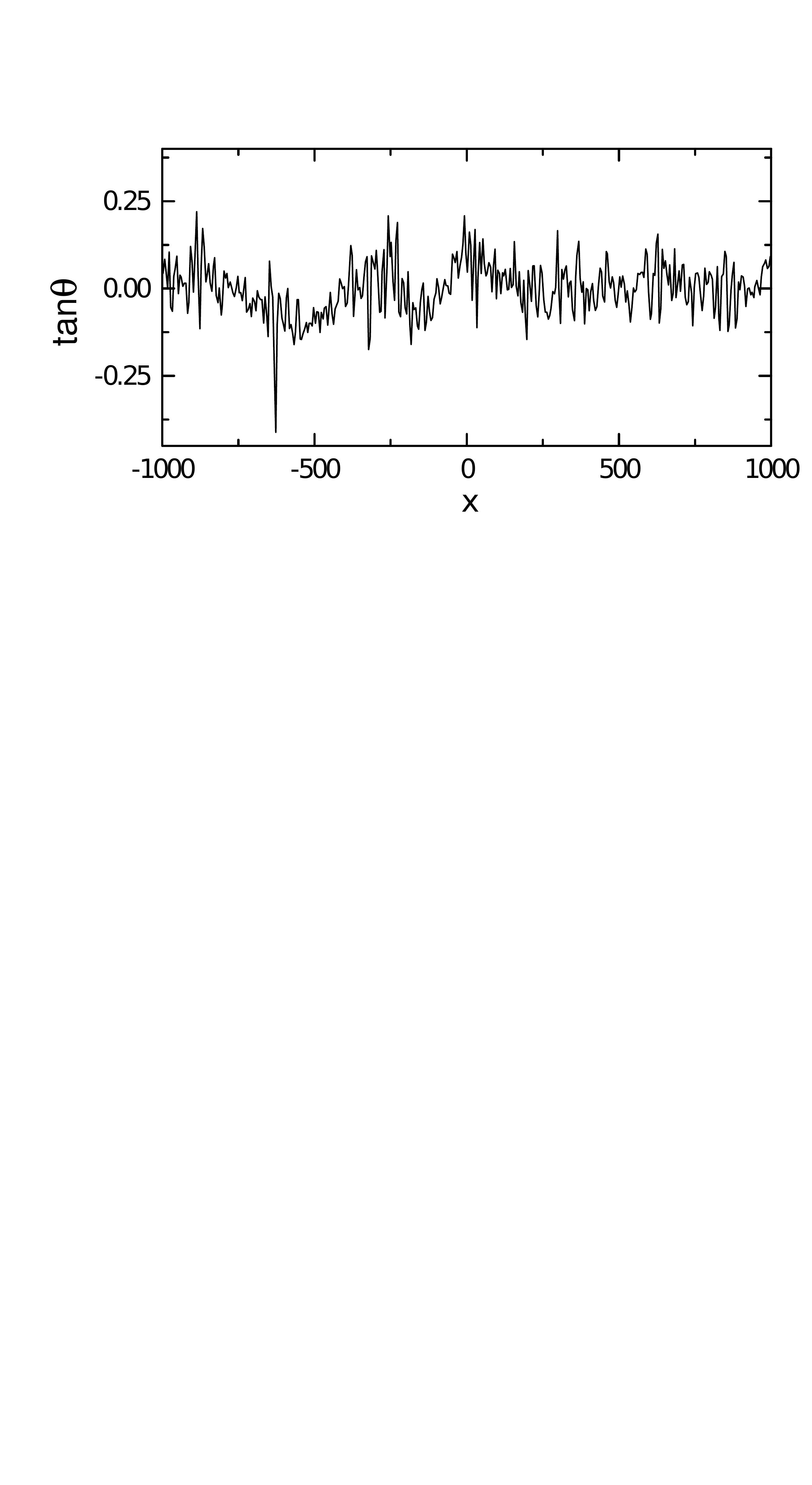}
		\caption{ Effective slope or $\tan{\theta}$ along the SB-path.
		}
		\label{fig:figure-power2}
	\end{center}
\end{figure*}

\begin{figure*}
	\begin{center}
		\includegraphics[width=1.4\columnwidth]{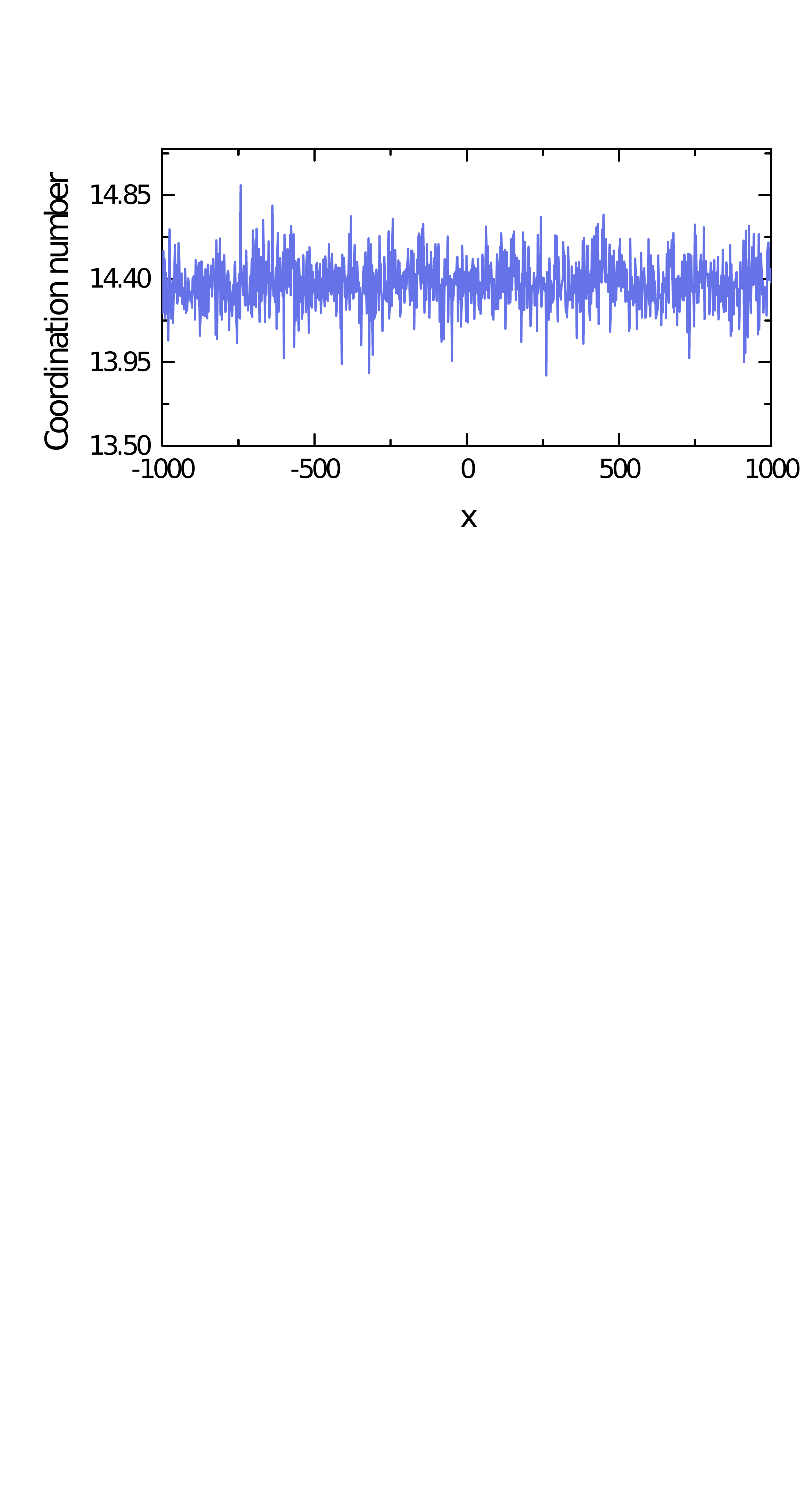}
		\caption{ Coordination number along the SB-path. 
		}
		\label{fig:figure-power3}
	\end{center}
\end{figure*}

\begin{figure*}
	\begin{center}
		\includegraphics[width=1.4\columnwidth]{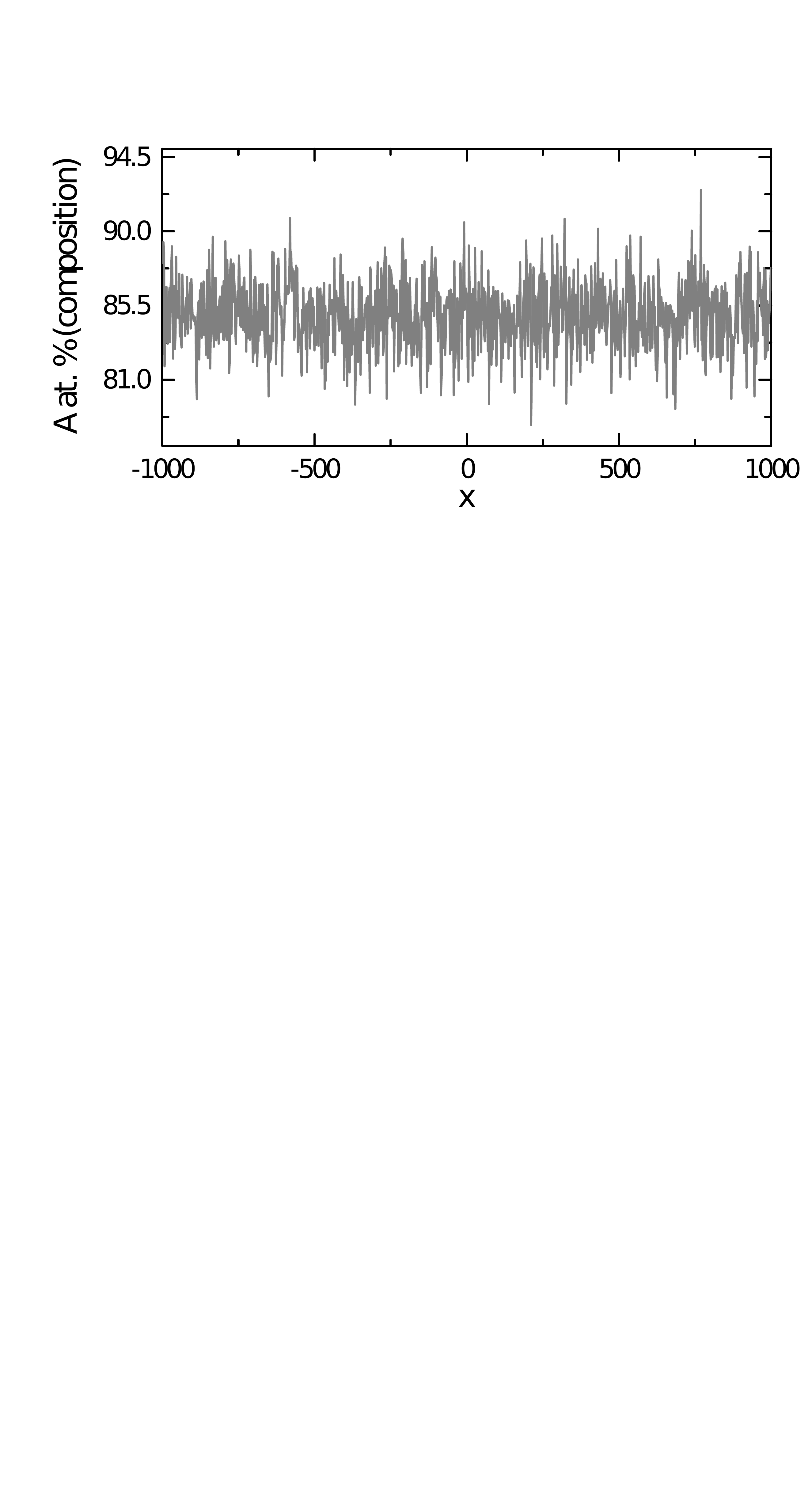}
		\caption{ Composition/
			percentage (\%) of A particles in $A_{x}B_{100-x}$ along the SB-path. 
		}
		\label{fig:figure-power4}
	\end{center}
\end{figure*}

\begin{figure*}
	\begin{center}
		\includegraphics[width=1.4\columnwidth]{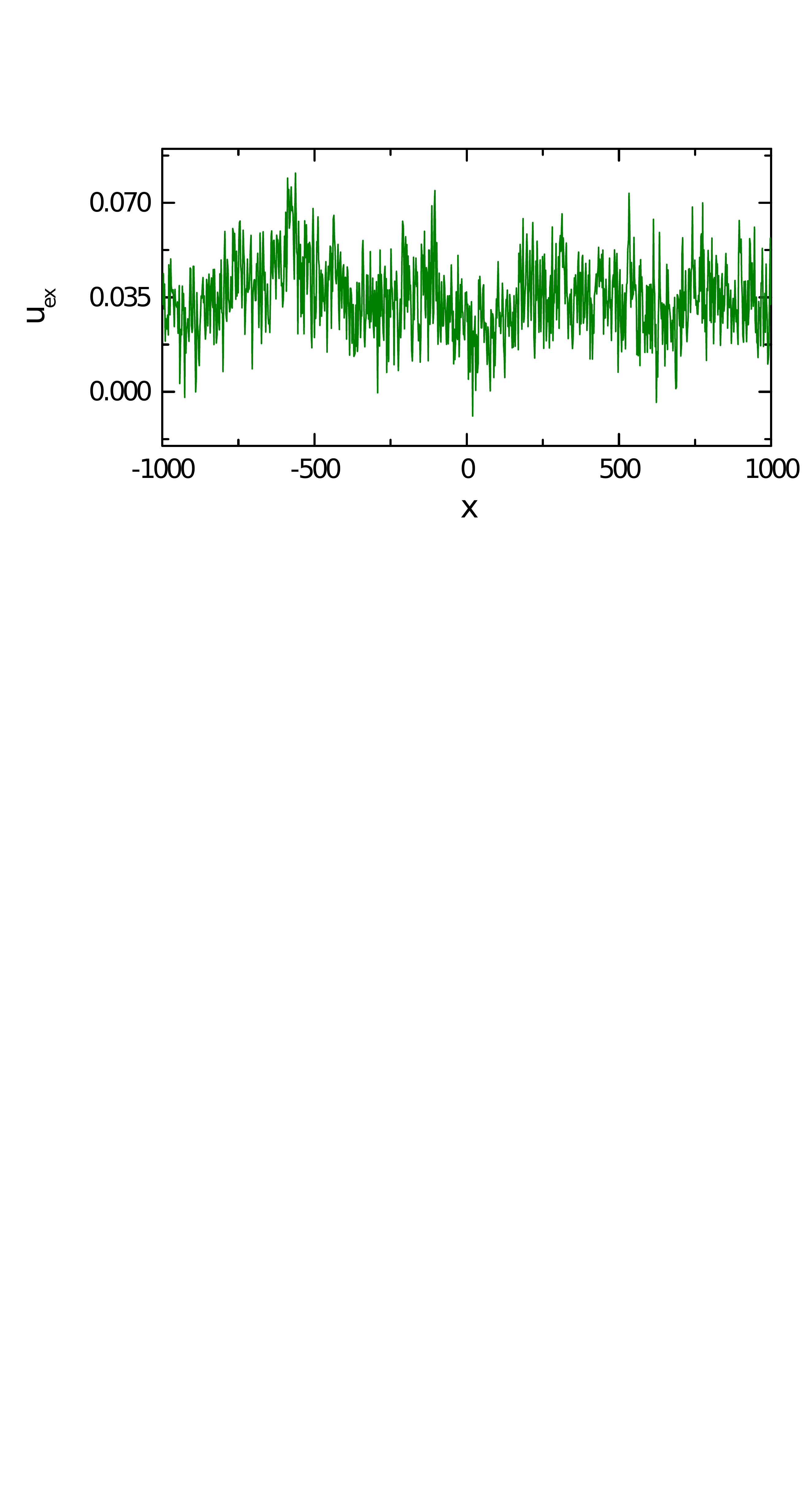}
		\caption{ Excess volume along the SB-path.  
		}
		\label{fig:figure-power5}
	\end{center}
\end{figure*}

\begin{figure*}
	\begin{center}
		\includegraphics[width=1.4\columnwidth]{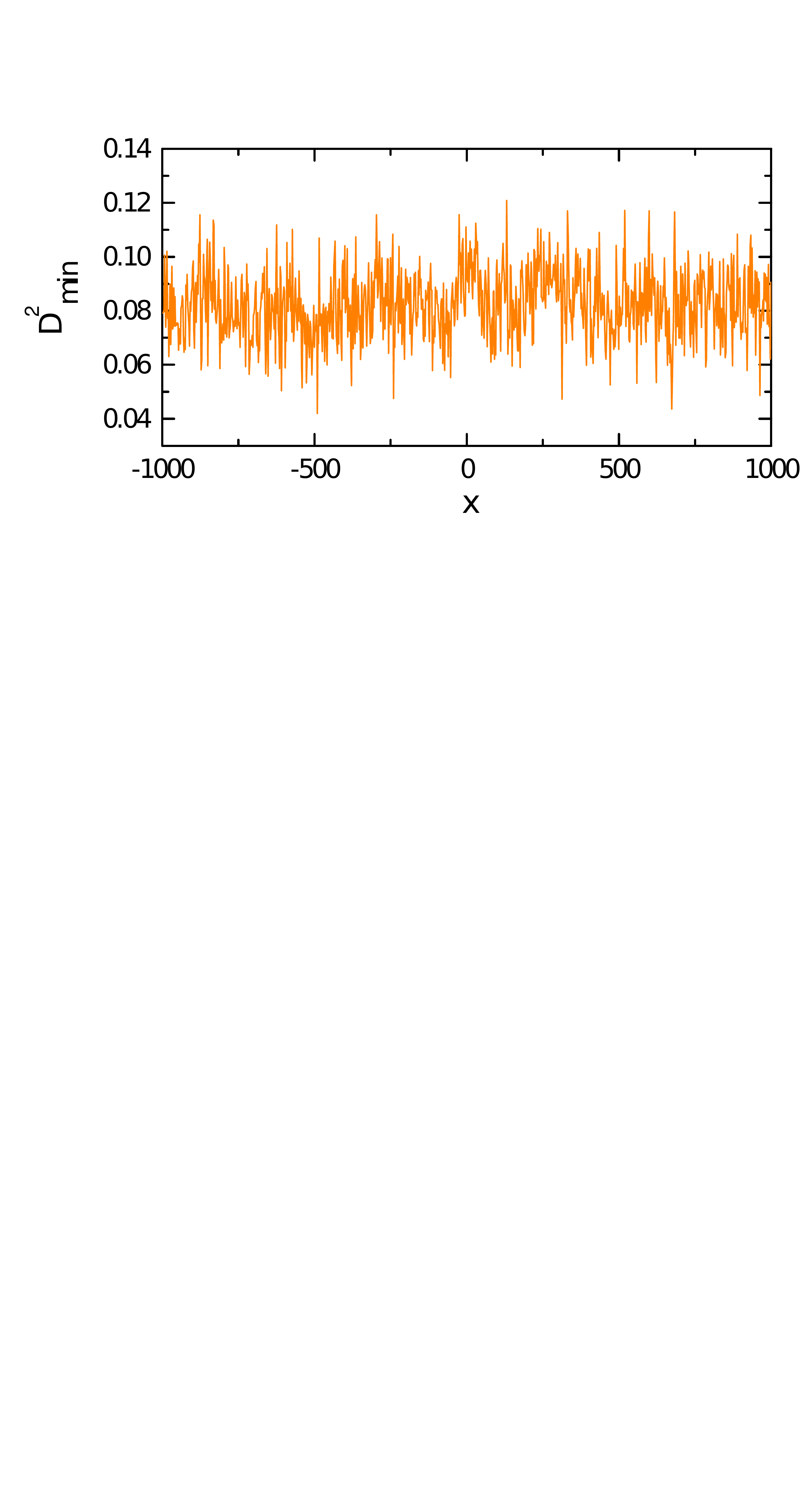}
		\caption{ Non affine deformation quantified by the $D^2_{min}$ parameter along the SB-path.
		}
		\label{fig:figure-power6}
	\end{center}
\end{figure*}

\begin{figure*}
	\begin{center}
		\includegraphics[width=1.4\columnwidth]{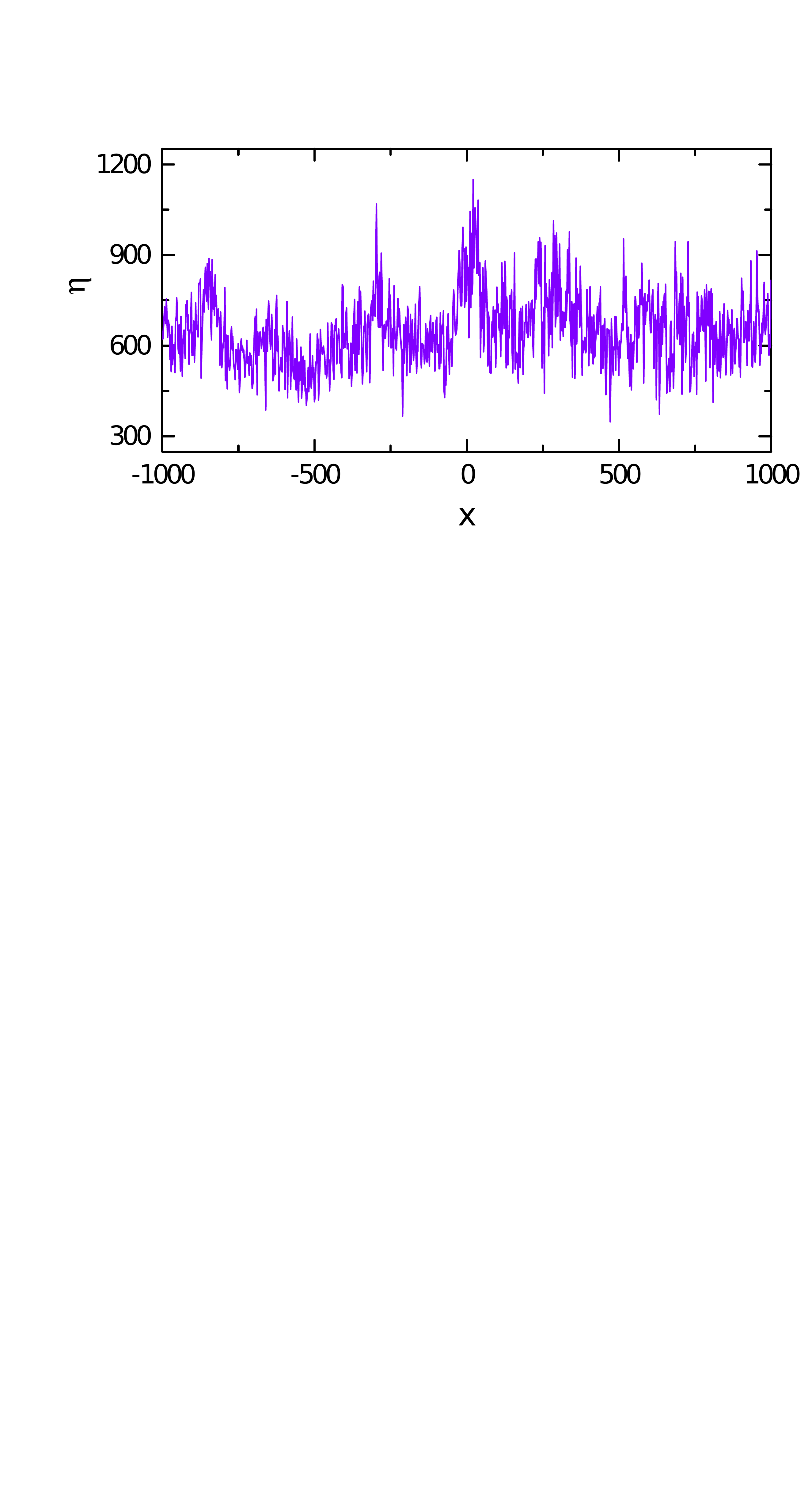}
		\caption{ Dynamic viscosity along the SB-path.  
		}
		\label{fig:figure-power7}
	\end{center}
\end{figure*}

\begin{figure*}
	\begin{center}
		\includegraphics[width=1.4\columnwidth]{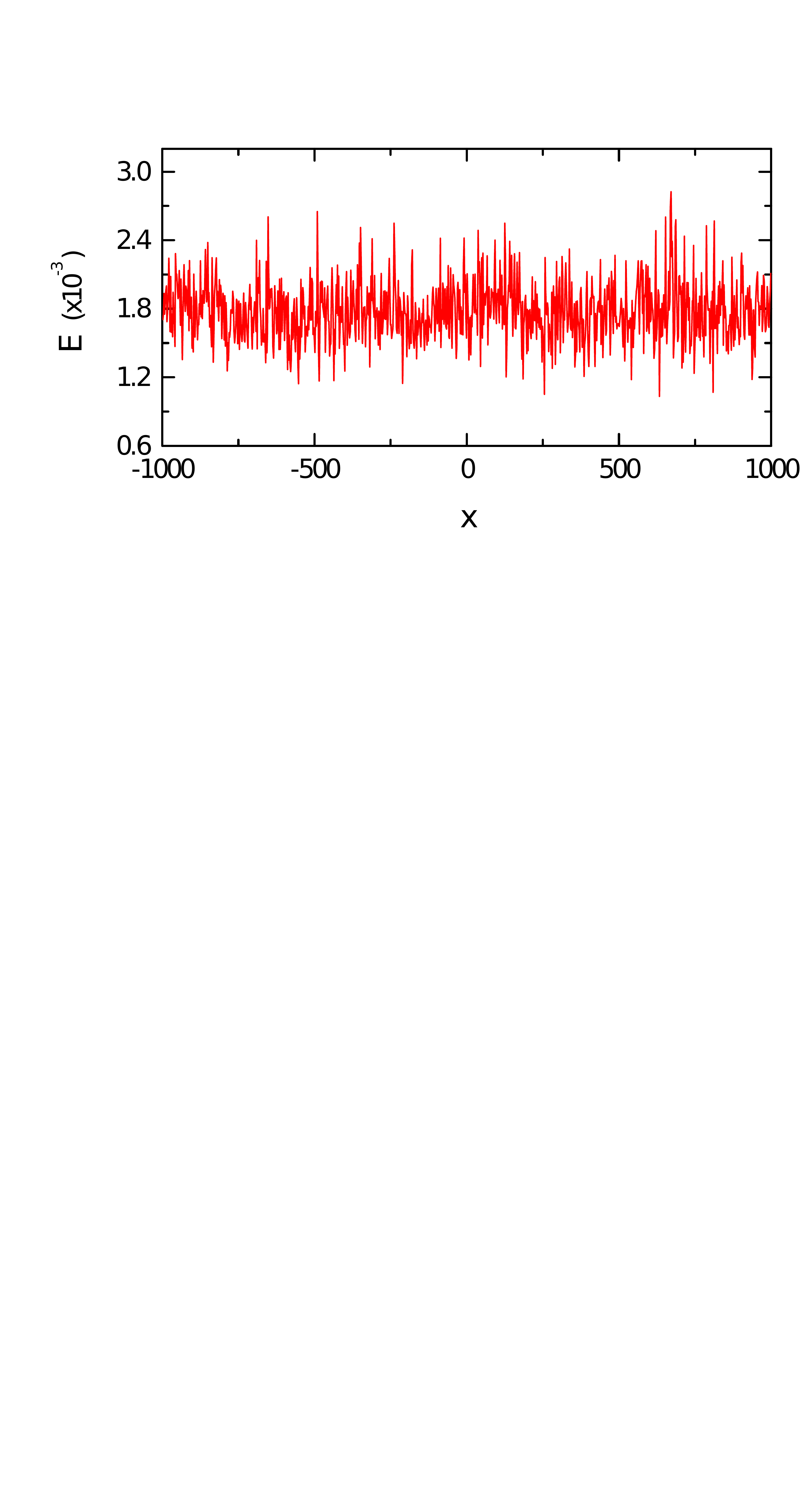}
		\caption{ Dissipated energy along the SB-path. 
		}
		\label{fig:figure-power8}
	\end{center}
\end{figure*}

\end{document}